\documentclass[amsmath,amssymb,onecolumn,superscriptaddress]{revtex4}

\usepackage{graphicx}%
\usepackage{dcolumn}%
\def\be{\begin{equation}}
\def\ee{\end{equation}}
\def\beq{\begin{eqnarray}}
\def\eeq{\end{eqnarray}}

\usepackage{bm}
\usepackage{graphicx, psfrag}
\usepackage[colorlinks=true, citecolor=blue, urlcolor = blue, linkcolor= red, bookmarks=true]{hyperref}
\usepackage{float}
\usepackage{amssymb}
\usepackage{amsmath}
\usepackage{hyperref}
\usepackage{subfigure}
\usepackage{pgfplots}
\usepackage{epstopdf}
\usepackage{booktabs}
\usepackage{orcidlink}
\usepackage{xcolor}

\begin{document}


\title{A study on matter accretion onto charged black hole solution in metric-affine gravity}  


\author{G. Mustafa}
\email[Email:]{gmustafa3828@gmail.com}
\affiliation{Department of Physics, Zhejiang Normal University, Jinhua 321004, People's Republic of China}

\author{A. Ditta}
\email[Email:]{adsmeerkhan@gmail.com}
\affiliation{Department of Mathematics, The Islamia University of Bahawalpur,
Bahawalpur-63100, Pakistan}

\author{Faisal Javed}
\email[Email:]{faisaljaved.math@gmail.com}
\affiliation{Department of
Physics, Zhejiang Normal University, Jinhua 321004, People's
Republic of China}

\author{S.K. Maurya}
\email[Email:]{sunil@unizwa.edu.om}
\affiliation{Department of Mathematical and Physical Sciences, College of Arts and Sciences, University of Nizwa, Nizwa 616, Sultanate of Oman}

 \author{Himanshu Chaudhary}
 \email[Email:]{himanshuch1729@gmail.com}
\affiliation{Department of Applied Mathematics, Delhi Technological University, Delhi-110042, India}
\affiliation{Pacif Institute of Cosmology and Selfology (PICS), Sagara, Sambalpur 768224, Odisha, India}
\affiliation{Department of Mathematics, Shyamlal College, University of Delhi, Delhi-110032, India,}

\author{Farruh Atamurotov}
\email[Email:]{atamurotov@yahoo.com}
\affiliation{Central Asian University, Milliy Bog' Street 264, Tashkent 111221, Uzbekistan}
\affiliation{University of Tashkent for Applied Sciences, Str. Gavhar 1, Tashkent 100149, Uzbekistan}
\affiliation{Institute of Theoretical Physics, National University of Uzbekistan, Tashkent 100174, Uzbekistan}

\date{\today}

\begin{abstract}
This study deals with astrophysical accretion onto the charged black hole solution, which is sourced by the dilation, spin, and shear charge of matter in metric affine gravity. The metric affine gravity defines the link between torsion and nonmetricity in space-time geometry. In the current analysis, we study the accretion process of various perfect fluids that are accreting near the charged black hole in the framework of metric affine gravity. Within the domain of accretion, multiple fluids have been examined depending on the value of $f_1$. The ultra-stiff, ultra-relativistic, and sub-relativistic fluids are considered to discuss the accretion. In the framework of equations of state, we consider isothermal fluids for this investigation. Further, we explore the effect of polytropic test fluid in relation to accretion discs, and it is presented in phase diagrams. Some important aspects of the accretion process are investigated. Analyzing the accretion rate close to a charged black hole solution, typical behavior is created and discussed graphically. \\
\textbf{keywords:} Black Hole; Hamiltonian approach; Accretion; Metric affine gravity.
\end{abstract}

\maketitle

\section{Introduction}

Modified Einstein gravity refers to theories that modify or extend Einstein's general relativity (GR). These modifications are introduced to address certain shortcomings or limitations of the original theory, such as the inability to explain the accelerated expansion of the universe or the existence of dark matter and dark energy. These modifications often involve introducing additional fields, modifying the equations of motion, or considering higher-dimensional spacetimes. The non-metricity of spacetime is explained by extra components introduced in modified Einstein field equations, such as metric affine gravity (MAG) \cite{2f}. The purpose of these changes is to provide a more complete account of gravity beyond GR by including the effects of torsion and non-metricity into the gravitational field equations. Metric affine gravity sheds light on the underlying characteristics of gravity and its interactions with matter and energy by taking into account geometric features of spacetime other than curvature.  

The sources of torsion in matter are specifically found to be the spin angular momentum \cite{3f,4f}, while the sources of nonmetricity are found to be the so called dilation and shear currents of matter \cite{6f,8f}. Development of new models in the field of MAG that provide precise solutions with torsion and nonmetricity beyond GR is especially important, since there are currently insufficient better understandings and empirical evidence for the existence of these variables. Unfortunately, the main obstacle to finding such models is the computational complexity resulting from the simultaneous inclusion of curvature, torsion, and nonmetricity in the computations. Furthermore, the order of the corresponding field strength tensors in the gravitational action determines the dynamics of these post-Riemannian quantities. For material sources with inherent hypermomentum \cite{9f}, the torsion and nonmetricity tensors are connected to non-propagating linear combinations of the curvature tensor. Conversely, the introduction of dynamics to these fields through higher-order curvature corrections results in a notable rise in theoretical complexity \cite{10f}. Despite these difficulties, from the beginning of MAG, a great deal of work has been done to comprehend the physical effects of the field equations and to identify and examine different kinds of solutions \cite{11f,12f,13f,14f,15f,16f,17f,18f,19f}. Notably, recent research has investigated the use of MAG in gravitational waves, black hole (BH) physics, and cosmology \cite{20f,21f,22f,23f,24f,25f,26f,27f,28f,30f,31f}.

The first BH solutions with torsion and nonmetricity were found by looking at a basic Lagrangian form and applying mathematical restrictions to these tensors. These limitations, which significantly simplify the field equations, include the triplet ansatz and other limits produced via prolongation techniques \cite{32f,33f}. These simplifications, however, lead to BH configurations that do not adequately display the dynamic link among the coframe, the metric, and the independent post-Riemannian structures found in the connection. A novel model was recently developed in the context of Weyl-Cartan geometry, integrating nonmetricity and dynamical torsion and establishing a relationship between these components \cite{29f}. Investigating the dynamic role of the traceless component of the nonmetricity tensor and its relation to shears is the main focus of study by Bahamonde and others \cite{34f}. They have created a model that takes this dynamic into account, leading to a novel BH solution with shear charges. This extension is shown to provide, upon the introduction of dynamical torsion and a Weyl vector, the largest known set of static and spherically symmetric BH solutions with spin, dilation, and shear charges in Metric-Affine Gravity \cite{34f}. Recently, the dynamical configurations of thin-shell developed from BHs in MAG filled with massless and massive scalar fields were studied in \cite{35f}. Also, the thermodynamics and Joule-Thomson expansion of charged anti-de Sitter (AdS) BHs in MAG are studied in \cite{36f}. A. Ditta et al. were investigated testing BH in MAG with its parameters using particle dynamics and photon motion in details~\cite{Farruh2022}.  

A BH accretion disc is a swirling disk of gas and dust that forms around a BH as it pulls in surrounding matter through its strong gravitational pull. This disc is often made up of hot, ionized gas that emits intense radiation as it spirals inward towards the BH's event horizon. The accretion disc plays a crucial role in understanding the dynamics and energy release mechanisms associated with BHs. Exploration of accurate and analytical answers has led to continuous advancements in accretion theory \cite{34d}. These answers provide a framework to define different astrophysical circumstances, which is essential to understanding the events under study. Moreover, because they act as benchmark tests for numerical codes, analytical solutions are useful tools in the area \cite{35d}. The Bondi model provides an explanation of the uniform flow of a spherically symmetric fluid falling onto a BH in the context of Newtonian gravity \cite{34d}. This model was expanded by Michel to the relativistic regime using a Schwarzschild BH as a framework  \cite{36d}. Furthermore, Tejeda and Aguayortiz produced analytic solutions for a Schwarzschild BH \cite{38d}, while Bondi and Hoyle created analytic solutions for wind accretion situations in the Newtonian theory\cite{37d}. Many studies using both analytical and numerical methods have improved our understanding of spherical and wind accretion \cite{40,41,42,43,44,45}. The studies of various BH spacetimes are explored in \cite{t1}-\cite{t15}.

Researchers have investigated phantom accretion inside a certain type of BHs \cite{46,47,48,49}. Furthermore, utilizing massless and massive scalar fields, a number of research publications have examined the dynamic configurations of particles at the boundary of different BH geometries \cite{50a,50bb,50b,50k}. In \cite{50,51,52}, Ditta and Abbas have additionally demonstrated efficacious outcomes concerning accretion while taking into account various kinds of BHs. In the setting of nonlinear electrodynamics BHs, Ditta et al. recently examined circular orbits and accretion disks \cite{rr52}. Furthermore, researchers have investigated modified theories of gravity, including noncommutative theory \cite{54}, quantum gravity corrections \cite{55}, and accretion flows onto scalar-tensor-vector gravity \cite{53}. It has also been explored \cite{56,57} for heteroclinic and cyclic accretion processes onto $f(R)$  and $f(T)$ BHs. Furthermore, studies have been conducted \cite{58}-\cite{63} utilizing dynamical systems and BH thermodynamics to analyze accretion flows onto static BHs. Researchers have also carried out fascinating research on alternative BH geometries, including \cite{p1}-\cite{p11}. Quasi-periodic oscillations were computed and tested by Liu and colleagues \cite{rr53,rr54} with observational data from various compact objects.

The settlement of this paper is as follows: section \textbf{2} is devoted to present BH solution in MAG. In section \textbf{3}, we determine the generic expressions of accretion of BH solutions in MAG. The speed of sound at the sonic points are evaluated in the section \textbf{4} by considering Hamiltonian system for the accretion by the equation of state (EoS). Moreover, in section \textbf{5}, we obsere the physical configurations of accretion flow near a considered BH in MAG for numerous cases of fluids such as ultra-stiff fluid (USF), ultra-relativistic fluid (URF), radiation fluid (RF) and sub-relativistic fluid (SRF). The polytropic test fluid accretion has been analyzed in section \textbf{6}. The accretion rate for polytropic fluid and mass accretion rate for different fluid distributions are presented in \textbf{7} and \textbf{8}, respectively. The last section presents the final outcomes of the our manuscript.

\section{Metric-Affine Gravity and black hole solution with spin, dilation and shear charges}
\begin{figure*}
\begin{center}
\includegraphics[width=8.3cm,height=5.5cm]{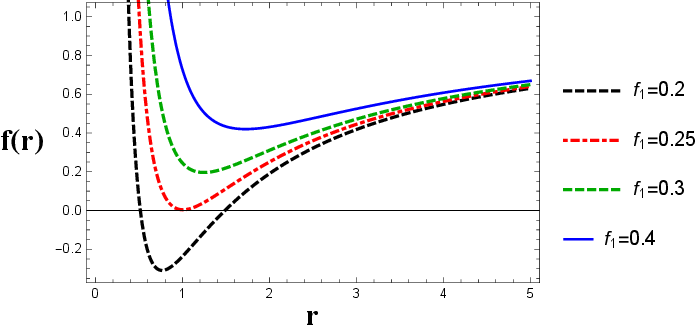}~~~
\includegraphics[width=8.3cm,height=5.5cm]{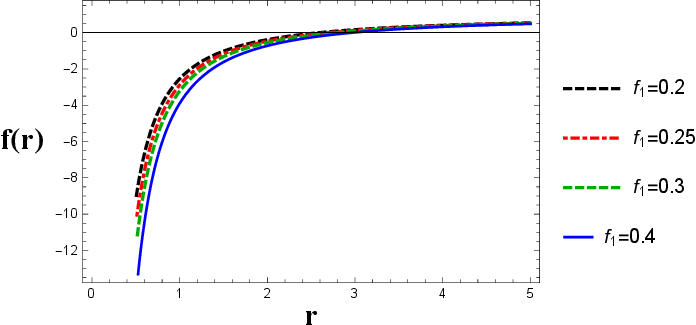}
\includegraphics[width=7cm,height=5.8cm]{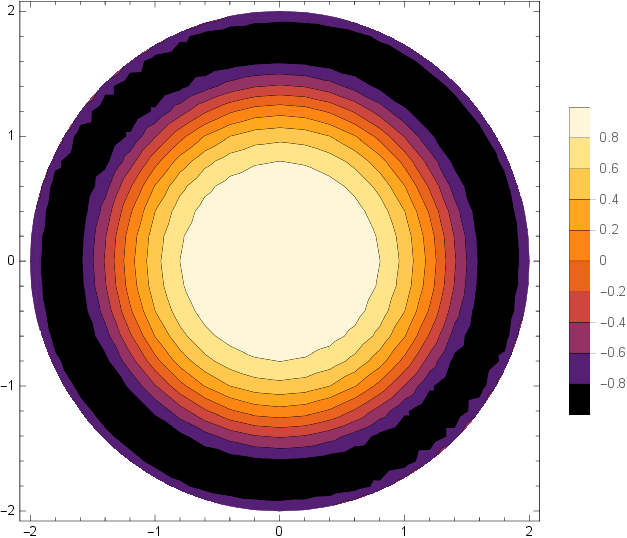}~~~~~~~~~~~~~~~~~~~~~
\includegraphics[width=7cm,height=5.8cm]{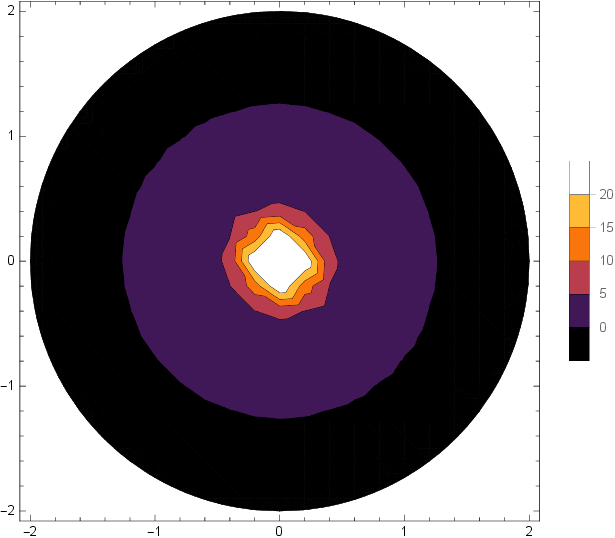}
\caption{Horizon display using the MAGBH parameters. Left plot is for $d_1=8f_1$ in the lapse function $f(r)=0$ with the variation in $f_1$.
Right plot is for $d_1=-8f_1$ in the lapse function $f(r)=0$ with the variation in $f_1$. The fixed parameters are $m=1$, $e_1=0.1$, $k_s=0.85$, $k_d=0.7$ and $k_{sh}=0.7$. The bottom plots are for mass of MAGBH with the conditions $d_1=8f_1$ and $d_1=-8f_1$.}\label{f1}
\end{center}
\end{figure*}

The MAG is a gravitational theory that expands beyond GR by accommodating non-zero torsion and nonmetricity. The literature around a non-holonomic connection $w_{\epsilon} \in g[(4,\mathrm{R})$ explains the concepts of the spin, dilation and shear currents of matter, and fulfils via a vector bundle isomorphism, or coframe field, which is further defined by the following relation $e^{a}_{epsilon}$. The relation of an affine-exhibiting metric spacetime is expressed as  \cite{34f}:
\begin{equation}\label{1}
    w^a{}_{b\epsilon}=e^a{}_{\lambda} e_b{}^{\rho}\tilde{\Gamma}^{\lambda}_{\rho\epsilon}+e^{a}{}_{\lambda}\partial_{\epsilon}e_{b}{}^{\lambda},
\end{equation}
which is fully quantified by the torsion and nonmetricity tensors as \cite{34f}
\begin{eqnarray*}
T^\lambda_{\epsilon\psi   } = 2\tilde{\Gamma}^{\lambda}_{[\epsilon\psi   ]}, \quad Q_{\lambda\epsilon\psi   } = \tilde{\nabla}_{\lambda} g{\epsilon\psi}.
\end{eqnarray*}
Let us start from the modified action in the form of dynamical nonmetricity tensor, which is provided as: \cite{34f}
\begin{small}
\begin{eqnarray} \label{2}
S=\int d^4 x \sqrt{-g}\left\{\mathcal{L}_{\mathrm{m}}+\frac{1}{16 \pi}\left[-R+2 f_1 \tilde{R}_{(\lambda \rho) \epsilon \psi   } \tilde{R}^{(\lambda \rho) \epsilon \psi   }+2 f_2\left(\tilde{R}_{(\epsilon \psi   )}-\hat{R}_{(\epsilon \psi   )}\right)\left(\tilde{R}^{(\epsilon \psi   )}-\hat{R}^{(\epsilon \psi   )}\right)\right]\right\},~
\end{eqnarray}
\end{small}
where $\mathcal{L}_{\mathrm{m}}$ represents the matter Lagrangian. The nonmetricity field in the action within the scope of symmetric part of the curvature tensor and its contraction is defined as \cite{34f}
\begin{eqnarray}\label{3}
\tilde{R}^{(\lambda \rho)}{ }_{\epsilon \psi   } =&&\tilde{\nabla}_{[\psi   } Q_{\epsilon]}{ }^{\lambda \rho}+\frac{1}{2} T_{\epsilon \psi   }^\sigma Q_\sigma^{\lambda \rho}, \nonumber\\
\tilde{R}_{(\epsilon \psi   )}-\hat{R}_{(\epsilon \psi   )} =&&\tilde{\nabla}_{(\epsilon} Q_{\psi   ) \lambda}^\lambda-\tilde{\nabla}_\lambda Q_{(\epsilon \psi   )}{ }^\lambda-Q^{\lambda \rho}{ }_\lambda Q_{(\epsilon \psi   ) \rho}+Q_{\lambda \rho(\epsilon} Q_{\psi   )}{ }^{\lambda \rho}+T_{\lambda \rho(\epsilon} Q_{\psi    \rho}^{\lambda \rho}.
\end{eqnarray}
Now, by varying the action by Eq. (\ref{2}) with respect to the anholonomic connection and co-frame field, one can find the following set of equations, which can be found in detail in the appendix part of Ref. \cite{34f}
\begin{eqnarray}
Y 1_\epsilon^\psi    =8 \pi \theta_\epsilon{ }^\psi   ,\label{4}\\
Y 2^{\lambda \epsilon \psi} =4 \pi \Delta^{\lambda \epsilon \psi   },\label{5}
\end{eqnarray}
where $Y 1_\epsilon{ }^\psi$ and $Y 2^{\lambda \epsilon \psi}$ provides the tensor quantities. Also, $\Delta^{\lambda \epsilon\psi   }$ and $\theta_\epsilon{ }^\psi   $ express hyper-momentum density tensor and the canonical energy-momentum tensor as
\begin{eqnarray}
\begin{aligned}\label{6}
\theta_\epsilon{ }^\psi    & =\frac{e^a{ }_\epsilon}{\sqrt{-g}} \frac{\delta\left(\mathcal{L}_m \sqrt{-g}\right)}{\delta e^a \psi   }, \\
\Delta^{\lambda \epsilon \psi} & =\frac{e^{a \lambda} e_{b \epsilon}}{\sqrt{-g}} \frac{\delta\left(\mathcal{L}_m \sqrt{-g}\right)}{\delta w^a{ }_{b \psi   }}.
\end{aligned}
\end{eqnarray}
For MAG, the anholonomic connection by Eq. (\ref{1}) in the Lie-algebra of the linear group $GL(4,\mathcal{R})$ separated into hyper-momentum showing its formal decay into spin, dilation, and shear currents \cite{rr60}. The necessary presence of shear gives rise to a dynamical traceless nonmetricity tensor within the composition of the special linear group $SL(4,\mathcal{R})\subset GL(4,\mathcal{R})$ \cite{rr61}. Finally, the effective gravitational action for these above quantities has the following form \cite{34f}:
\begin{small}
\begin{align}\label{7}
S= \frac{1}{64 \pi} \int d^4 x \sqrt{-g}\left[-4 R-6 d_1 \bar{R}_{\lambda[\rho \epsilon \psi   ]} \bar{R}^{\lambda[\rho \epsilon \psi   ]}-9 d_1 \bar{R}_{\lambda[\rho \epsilon \psi   ]} \bar{R}^{\epsilon[\lambda \psi    \rho]}+8 d_1 \bar{R}_{[\epsilon \psi   ]} \bar{R}^{[\epsilon \psi   ]}+4 e_1 \tilde{R}^\lambda{ }_{\lambda \epsilon \psi   } \tilde{R}^\rho{ }_\rho \epsilon \psi   \right. \nonumber\\
+8 f_1 \tilde{R}_{(\lambda \rho) \epsilon \psi   } \tilde{R}^{(\lambda \rho) \epsilon \psi   }-2 f_1\left(\tilde{R}_{(\epsilon \psi   )}-\hat{R}_{(\epsilon \psi   )}\right)\left(\tilde{R}^{(\epsilon \psi)}-\hat{R}^{(\epsilon \psi)}\right)+18 d_1 \bar{R}_{\epsilon[\lambda \rho \psi   ]} \tilde{R}^{(\epsilon \psi   ) \lambda \rho}-6 d_1 \bar{R}_{[\epsilon \psi   ]} \bar{R}^{\epsilon \psi } \nonumber\\
\left.-3 d_1 \tilde{R}_{(\lambda \rho) \epsilon \psi   } \tilde{R}^{(\lambda \rho) \epsilon \psi   }+6 d_1 \tilde{R}_{(\lambda \rho) \epsilon \psi } \tilde{R}^{(\lambda \epsilon) \rho \psi}+\frac{9}{2} d_1 \tilde{R}_{\epsilon \psi} \tilde{R}^{\epsilon \psi   }+3\left(1-2 a_2\right) T_{[\lambda \epsilon \psi   ]} T^{[\lambda \epsilon \psi   ]}\right],
    \end{align}
\end{small}
which provides the following independent field equations (the main calculations can be seen in appendix of Ref. \cite{34f}):
\begin{eqnarray}
      X1_{\epsilon}^{\psi   }=0,\label{8}\\
    X2^{\lambda\epsilon\psi   }=0,\label{9}
\end{eqnarray}
where $ X2^{\lambda\epsilon\psi}$ and $X1_{\epsilon}^{\psi}$ express the tensor quantities \cite{34f}. As a result, the tetrad field equations (\ref{8}-\ref{9}) help us to define the Reissner-Nordstr$\ddot{o}$m-like BH geometry with spin, dilation, and shear charges. Finally, we have the following lapse function, i.e., $g_{tt}$ metric component for Reissner-Nordstr$\ddot{o}$m-like BH geometry as \cite{34f,35f},
\begin{eqnarray}\label{10}
    f(r)&=&1-\frac{2 m}{r}+\frac{d_1 \kappa_{\rm s}^2-4 e_1 \kappa_{\rm d}^2-2f_1 \kappa_{\rm sh}^2}{r^2}.
\end{eqnarray}
In the metric function $f(r)$, the parametric representation is as follows:
$d_1$ and $f_1$ denote the Lagrangian coefficients; $e_1$ is a
constant whereas $k_{sh}$, $k_s$ and $k_d$ are the shear, spin and dilation charges, respectively. 
In this setup, one can produce three kinds of solutions, namely: 
(i) $d_1 = 8f_1$, (ii) $d_1=-8f_1$ and (iii) $d_1\neq8f_1$ for which $f_1 \leq 0$.
The conventional Reissner-Nordstr$\ddot{o}$m solution of General Relativity may be derived from this geometric framework. Instead of assuming an electromagnetic component, the solution is constructed using a traceless nonmetricity field in the given context.  To study the accretion properties of BHs, 
we just consider two cases of solutions, that is, $d_1 = 8f_1$ and $d_1=-8f_1$. 
The horizon radii for the metric function (\ref{2}) can be found as
\begin{equation}
r_\pm=M-\sqrt{4e_1k^2_d-d_1k^2_s+2f_1k^2_{sh}+M^2}.\label{7x}
\end{equation}
For the horizon analysis of the metric-affine gravity black hole (MAGBH), we plot $f(r)=0$ by varying the values of $f_1$ for both cases $d_1=\pm8f_1$ in Fig.~\ref{f1},
which behaves differently. The effect of $f_1$ is important and makes a constant contribution to the horizon analysis.
Therefore, the MAGBH horizon has the following structure:
As seen in Fig.~\ref{f1} (left plot),
\begin{itemize}
  \item We observe that an event horizon is formed by the effected parameter $f_1$ of MAGBH at $f_1=0.25$ (red dotes).
  \item Two horizons are formed at the initial value of he MAGBH parameter $f_1=0.2$ (black dotes).
  \item The other two curves have no horizon.
  \item Subsequently, the curves go away from the BH as the parameter $f_1$ increases.
  \end{itemize}

As seen in Figure (\ref{f1}), the right plot exhibits different behaviour compared to the left plot while varying the values of $f_1$. The maximum portion of the curve is negative. In this portion, the radius increased by increasing $f_1$ and vice versa. The bottom plots are significant because they support our observation of the light emitted by accreting MAGBH. This understanding can be used to characterize the nature of accretion and constrain the parameters of BH.

\section{Spherical accretion flow and some Fundamental equations}

In this section, our aim is to find the general solutions of accretion onto MAGBH, which explain the motion of particles around the horizon. To do this, we are assuming the basic conservation laws in the context of the spherically symmetric metric. During the accretion process, there is no mass annihilation, and creation leads to the law of mass conservation, which is expressed as
\begin{equation}
\nabla_\eta J^\eta =0.\label{08}
\end{equation}
As, $J^\eta=\varepsilon u^\eta$ in which $\varepsilon$ denotes the number of particles and
$u^\eta=\frac{dx^\eta}{d\tau}$ is the velocity of the fluid. Now, by introducing the energy-momentum tensor $T^{\eta\zeta} =(\rho+p)u^\eta u^\zeta+pg^{\eta\zeta}$, we can setup the current problem. Further, the expression $u^\eta=\frac{dx^\eta}{d\tau}=(u^t,u^r,0,0)$, is the four velocity of the particle used in the conservation of energy, which is provided as
\begin{equation}
\nabla_\eta T^{\eta\zeta} =0.\label{09}
\end{equation}
By applying the normalization condition $ u^\mu u_\mu = -1$, it is easy to find a required relation, which is defined as
\begin{equation}
\Big[u^t\Big]^2 =\frac{1}{f^2(r)}\Big[u^2+ f(r)\Big],\label{010}
\end{equation}
also
\begin{equation}
\Big[u^t\Big]^2 =u^2-f(r) .\label{11}
\end{equation}
For equatorial plane, we assume $(\theta=\pi/2)$, the mass conservation law by Eq. (\ref{08}) can be rewritten as
\begin{equation}
\Big[\frac{1}{r}\Big]^2\frac{d}{dr}\Big[\varepsilon ur^2\Big]=0.\label{12}
\end{equation}
Now, by integrating, we get the following expression
\begin{equation}
 \varepsilon ur^2=A_1,\label{13}
\end{equation}
where $A_1$ is integration constant. Since accretion belongs to the fluid velocity $u^r=u<0$, we define the enthalpy, which is the ratio between the density of the fluid and the total energy of its structure at a constant pressure (as $h(\rho,p,\varepsilon)=\frac{\rho+p}{\varepsilon}$). Since the flow of fluid is smooth, therefore Eq. (\ref{9}) yields
\begin{equation}
\varepsilon u^\eta\nabla_\eta (hu^\zeta)+g^{\eta\zeta} \partial_\eta p=0.\label{14}
\end{equation}
Next, if the entropy of a moving fluid along a streamline is constant, then the fluid flow must be isotropic \cite{42}. The above equation gives the following relationship: 
\begin{equation}
u^\eta \nabla_\eta (hu^\eta)+\partial_\zeta h=0.\label{15}
\end{equation}
Now, by putting the zeroth component, we obtain
\begin{equation}
\partial_r (hu_t)=0.\label{16}
\end{equation}
On integrating, one can get the following expression
\begin{equation}
\Big[h^2(f(r)+(u)^2)\Big]^{1/2}=A_2,\label{17}
\end{equation}
where $A_2$ is integration constant. These fundamental relations by Eq. (\ref{13}) and Eq. (\ref{17})
exploring the critical flow of fluid in the framework of spherically symmetric MAGBH.

\section{Dynamical Analysis}
In this current section, we shall discuss about sonic points and isothermal test fluid. 

\subsection{Sonic Points}
The dark compact object achieves maximum accretion when the flow passes through the sonic point.
This is a critical point where the velocity of falling matter is equal to the speed of sound. 
To determine these points under MAGBH, we consider barotropic matter with constant enthalpy as $h=h(\varepsilon)$. The EoS for barotropic matter is given by~\cite{56}

\begin{eqnarray}
\frac{1}{h}dh = \mathbf{s}^2\frac{d\varepsilon}{\varepsilon},\label{18}
\end{eqnarray}
where $\mathbf{s}$ is the speed of sound, and the EoS becomes $\ln h=\mathbf{s}^2\ln \varepsilon$.
Using Eqs. (\ref{10}), (\ref{17}) and (\ref{18}), we have the following relation
\begin{equation}
\left[\left(\frac{u}{u_t}\right)^2-\mathbf{s}^2\right](\ln u)_{,r}=\frac{1}{r(u_t)^2}\left[2\mathbf{s}^2(u_t)^2-\frac{1}{2}r f'(r)\right].\label{19}
\end{equation}
To determine the value of the speed of sound at the sonic point, we take both sides of the above Eq. (\ref{19}) equal to zero, which leads us 
\begin{equation}
\mathbf{s}^2_c = \left(\frac{u_c}{u_{tc}}\right)^2.\label{20}
\end{equation}
From the right hand side of Eq. (\ref{19}), we have
\begin{equation}
2\mathbf{s}^2_c(u_{tc})^2-\frac{1}{2}r_cf'_{r_c} = 0.\label{21}
\end{equation}
Using (\ref{20}) and (\ref{21}), attaining the critical radial velocity, given by the following relation
\begin{equation}
(u_c)^2 = \frac{1}{4}r_cf'_{r_c}.\label{22}
\end{equation}
Further, from Eqs. (\ref{11}), (\ref{21}) and (\ref{22}), one can get following expression
\begin{equation}
r_cf'_{r_c} = 4\mathbf{s}^2_c[f(r_c)+(u_c)^2].\label{23}
\end{equation}
Hence, the final value of the speed of sound is calculated as
\begin{equation}
\mathbf{\mathbf{s}}^2_c = \frac{r_cf'_{r_c}}{r_cf'_{r_c}+4f(r_c)}.\label{24}
\end{equation}
The Eqs. (\ref{22}) and (\ref{24}) provide the sonic points as $(r_c,\pm u_c)$, when we have the value of speed of sound.

\subsection{Isothermal Test Fluid Background}
The isothermal test fluid is a significant process which provides us a better understanding to explain the accretion flow for MAGBH at a constant temperature. Throughout the process, the speed of matter remains very fast, and it is exposed as adiabatic. In this process, considering $\mathbf{s}^2=\frac{dp}{d\rho}$, to determine analytic solutions of the equations of moving matter.
Now, by introducing the EoS $p=\omega\rho$, for these solutions, where $\rho$ is the energy density and $\omega$ is the state parameter with conditions $(0<\omega\leq1)$, defined in \cite{61rr}. By using $p=\omega\rho$ and $\mathbf{s}^2=\frac{dp}{d\rho}$, one can obtain $\mathbf{s}^2=\omega$. According to the first law of thermodynamics, we have 
\begin{equation}
 \frac{1}{dn}d\rho=\frac{\rho+p}{\varepsilon}=h.\label{25}
\end{equation}
By limit of integration from the sonic point to any point of the fluid, which is inside, is calculated as
\begin{equation}
\frac{\varepsilon}{\varepsilon_c}= \exp\left(\int^{\rho}_{\rho_c}\frac{1}{\rho'+p(\rho')}d\rho'\right).\label{26}
\end{equation}
By applying $p=\omega\rho$, the Eq. (\ref{26}) becomes
\begin{equation}
\frac{\varepsilon}{\varepsilon_c}=\left(\frac{\rho}{\rho_c}\right)^\frac{1}{\omega+1}.\label{27}
\end{equation}
By taking enthalpy as $h(\rho,p,\varepsilon)=\frac{\rho+p}{\varepsilon}$, into Eq. (\ref{27}), one can obtain a following relation
\begin{eqnarray}
\frac{(\omega+1)\rho_c}{\varepsilon_c}\left(\frac{\varepsilon}{\varepsilon_c}\right)^\omega=h.\label{28}
\end{eqnarray}
Now, by putting $A_3=\frac{A_2\varepsilon^{1-\omega}_c}{(\omega+1)\rho_c}$ into Eq. (\ref{28}) and Eq.(\ref{17}), we obtain the following result
\begin{equation}
\varepsilon^\omega\sqrt{f(r)+(u)^2}=A_3.\label{29}
\end{equation}
The Eq. (\ref{25}) and Eq. (\ref{10}), lead us to
\begin{equation}
\Big(f(r)+(u)^2\Big)^{1/2}=A_3r^{2\omega}(u)^\omega.\label{30}
\end{equation}
Therefore, we take the following Hamiltonian, which is defined in \cite{42,43}
\begin{equation}
H=\frac{f^{1-\omega}(r)}{(1-v^2)^{1-\omega}v^{2\omega}r^{4\omega}},\label{31}
\end{equation}
where $v \equiv \frac{dr}{f(r)dt}$, is three-dimensional speed of a particle with the
radial motion in equatorial plane, and it is defined as
\begin{equation}
v^2=\left(\frac{u}{f(r)u^0}\right)^2=\frac{u^2}{u^2_0}=\frac{u^2}{f(r)+u^2}.\label{32}
\end{equation}
The main critical solutions are explored at the sonic points, from Eqs. (\ref{22}) and (\ref{23}), which is given as
\begin{eqnarray}
(u_c)^2&=&\frac{1}{4}r_c f'_c(r),\label{33}\\
(u_c)^2 &=&\omega\left(f_c(r)+\frac{1}{4}r_c f'_c(r)\right).\label{34}
\end{eqnarray}
This generalized result can be analyzed by taking the any value of $\omega$.
In this work, we will use some interesting fluids such as UST ($\omega=1$), URF ($\omega=1/2$),
RF ($\omega=1/3$) and SRF ($\omega=1/4$) in the accretion flow around a MAGBH.

\section{Testing of Fluids }
The current section of this study deals with the Hamiltonian of different kinds of fluids, say USF, URF, RF, and SRF.

\subsection{The Hamiltonian of USF}

One can be found that pressure is equal to the energy density that is $p=\rho$ for the USF by using the EoS ($\omega=1$).
The critical radius $r_c$ is equal to the horizon that is $r_h=r_c$, at the metric potential $f_c=0$.
The Hamiltonian (\ref{31}) for this case is given by
\begin{equation}
H=\Big(\frac{1}{\nu r^2}\Big)^2.\label{36}
\end{equation}

\subsection{The Hamiltonian of URF}
It can be seen that the energy density in the URF is greater than the pressure, where the pressure is given by the equation $p=\rho/2$ and the parameter $\omega$ is equal to $1/2$. By the relation $r_c f'_c(r)-4f_c(r)=0$, the critical radius for URF may be derived by solving the Eqs. (\ref{33}) and (\ref{34}).
\begin{equation}
\frac{-6d_1k^2_s+12f_1k^2_{sh}+10mr-4r^2+24e_1k^2_d}{r^2}=0.\label{37}
\end{equation}
So, we choose only useful numerical value by the parameters $d_1=8f_1$, $f_1=1$, $e_1=1$, $m=1$ and $k_s=1$, $k_d=0.7$ $k_{sh}=2$.
\begin{equation}
r_c=3.37191.\label{38}
\end{equation}
In second condition $d_1=-8f_1$, the critical radius is $r_c=6.58877$.
Using the critical radius $r_c$, into (\ref{32}), we get another critical value $v_c$.
Then, by using the critical values $(r_c,\pm {v_c})$, with the result $s^2_c=v^2_c$, the critical Hamiltonian $H_c$
is obtained. The general Hamiltonian (\ref{31}) reduces into the form
\begin{equation}
H=\frac{f(r))^{1/2}}{r^2v (1-v^2)^{1/2}}.\label{39}
\end{equation}
This implies that
\begin{equation}
H=\frac{\left(1-\frac{2m}{r}+\frac{d_1k^2_s-4e_1k^2_d-2f_1k^2_{sh}}{r^2}\right)^{1/2}}{r^2v (1-v^2)^{1/2}}.\label{39x}
\end{equation}
By choosing the suitable value of $H=H_c$, one can see the graphical analysis of $v$ and $r_c$.
Continuing from (\ref{39}), we have 
\begin{equation}
v^2=\frac{1\pm \Big(1-4x(r)\Big)^{1/2}}{2},\label{40}
\end{equation}
where $x(r)=\frac{\Big(1-\frac{2MG}{r}+\frac{Q^2}{2\alpha r^2}-\frac{(1-\alpha)}{24\pi G c_0}r^2\Big)}{Hr^4}$.

\begin{table*}
    \caption{The sonic points for URF with respect to MAGBH parameter $f_1$}\label{tab1}
    \centering 
  \scalebox{0.85}{  \begin{tabular}{c c c c c c c c c c } 
        \hline 
                                      ~~~~~~~~~~~~~&$d_1=8f_1$~~~   \\ [0.1ex]
          \hline 
    $f_1$& & $r_{c}$ &~~~~~~~~~~~~~ &$v_{c}$ &  ~~~~~~~~~~~~~~~~~~~&~$H_{c}$ &~~~~~~ & \\ [0.1ex]
        \hline 
         $0.10$ ~~~~&~~   &$3.83118$&  ~~~~&~~  $0.197643$& ~~~~~&~~       $0.174527$&~~~~~~~~~~~~   \\
         $0.20$ ~~~~&~~   &$3.65052$&  ~~~~&~~  $0.262114$& ~~~~&~~         $0.1504$&~~~~~~~~~~~     \\
         $0.30$ ~~~~&~~   &$3.15069$&  ~~~~&~~  $0.21385$& ~~~~&~~        $0.19167$&~~~~~~~~~~   \\
         $0.40$ ~~~~&~~   &$2.49156$&  ~~~~&~~  $0.15296$& ~~~~&~~        $0.29775$&~~~~~~~~~~   \\
         $0.50$ ~~~~&~~   &$2.61044$&  ~~~~&~~  $0.16219$& ~~~~&~~        $0.33483$&~~~~~~~~~~~   \\
         $0.60$ ~~~~&~~   &$2.72054$&  ~~~~&~~  $0.174106$& ~~~~&~~        $0.361829$&~~~~~~~~~~~  \\
    \hline 
    \end{tabular}}
\end{table*}

\begin{table*}
    \caption{The sonic points for URF with respect to MAGBH parameter $f_1$}\label{tab2}
    \centering 
  \scalebox{0.85}{  \begin{tabular}{c c c c c c c c c c }
        \hline 
                                      ~~~~~~~~~~~~~&$d_1=-8f_1$~~~~~~~   \\ [0.5ex]
          \hline 
    $f_1$& & $r_{c}$ &~~~~~~~~~~~~~ &$v_{c}$ &  ~~~~~~~~~~~~~~~~~~~&~$H_{c}$ &~~~~~~ & \\ [0.5ex]
         \hline 
         $0.10$ ~~~~&~~   &$4.2604$&  ~~~~&~~  $0.203247$& ~~~~~&~~       $0.139826$&~~~~~~~~~~~~   \\
         $0.20$ ~~~~&~~   &$3.1308$&  ~~~~&~~  $0.235270$& ~~~~&~~         $0.14999$&~~~~~~~~~~~     \\
         $0.30$ ~~~~&~~   &$2.5966$&  ~~~~&~~  $0.20587$& ~~~~&~~        $0.18544$&~~~~~~~~~~   \\
         $0.40$ ~~~~&~~   &$2.19952$&  ~~~~&~~  $0.187071$& ~~~~&~~        $0.24782$&~~~~~~~~~~   \\
         $0.50$ ~~~~&~~   &$1.68072$&  ~~~~&~~  $0.15573$& ~~~~&~~        $0.29761$&~~~~~~~~~~~   \\
         $0.60$ ~~~~&~~   &$1.32235$&  ~~~~&~~  $0.24478$& ~~~~&~~        $0.321652$&~~~~~~~~~~~  \\
    \hline 
    \end{tabular}}
\end{table*}

\subsection{The Hamiltonian of RF}
One can find that the energy density is greater than three times of isotropic pressure that is $(3p=\rho)$ in the radiation fluid.
By the relation $r_c f'_c(r)-2f_c(r)=0$, the critical equation is given by
\begin{equation}
-\frac{2\Big(2d_1k^2_s-4f_1k^2_{sh}-3mr-r^2-8e_1k^2_d\Big)}{r^2}=0.\label{41}
\end{equation}
So, we choose only useful numerical values by the parameters $d_1=8f_1$, $f_1=1$, $e_1=1$, $m=1$ and $k_s=1$, $k_d=0.7$ $k_{sh}=2$.
\begin{equation}
r_c=3.98395.\label{42}
\end{equation}
In the second condition $d_1=-8f_1$, the critical radius is $r_c=7.67819$.
For this case, the general Hamiltonian (\ref{31}) is given by
\begin{equation}
H=\frac{f(r)^{2/3}}{r^\frac{4}{3}v^{\frac{2}{3}}(1-v^{2})^{\frac{2}{3}}}.\label{43}
\end{equation}
This implies that
\begin{equation}
H=\frac{\left(1-\frac{2m}{r}+\frac{d_1k^2_s-4e_1k^2_d-2f_1k^2_{sh}}{r^2}\right)^{\frac{2}{3}}}{r^\frac{4}{3}v^{\frac{2}{3}}(1-v^{2})^{\frac{2}{3}}}.\label{43x}
\end{equation}

\begin{table*}
    \caption{The sonic points for RF with respect to MAGBH parameter $f_1$}\label{tab3} 
    \centering 
   \scalebox{0.85}{ \begin{tabular}{c c c c c c c c c c }
        \hline 
        ~~~~~~~~~~~~~&$d_1=8f_1$~~~~~~~   \\ [0.5ex]
          \hline 
    $f_1$& & $r_{c}$ &~~~~~~~~~~~~~ &$v_{c}$ &  ~~~~~~~~~~~~~~~~~~~&~$H_{c}$ &~~~~~~ & \\ [0.5ex]
    \hline 
         $0.10$ ~~~~&~~   &$4.50832$&  ~~~~&~~  $0.0886502$& ~~~~~&~~       $0.557103$&~~~~~~~~~~~~   \\
         $0.20$ ~~~~&~~   &$4.24872$&  ~~~~&~~  $0.0866521$& ~~~~&~~         $0.55996$&~~~~~~~~~~~     \\
         $0.30$ ~~~~&~~   &$3.87663$&  ~~~~&~~  $0.0786564$& ~~~~&~~        $0.567354$&~~~~~~~~~~   \\
         $0.40$ ~~~~&~~   &$3.29723$&  ~~~~&~~  $0.182596$& ~~~~&~~        $0.58534$&~~~~~~~~~~   \\
         $0.50$ ~~~~&~~   &$2.18278$&  ~~~~&~~  $0.186002$& ~~~~&~~        $0.64298$&~~~~~~~~~~~   \\
         $0.60$ ~~~~&~~   &$1.99231$&  ~~~~&~~  $0.177754$& ~~~~&~~        $0.757143$&~~~~~~~~~~~  \\
    \hline 
    \end{tabular}}
\end{table*}

\begin{table*}
    \caption{The sonic points for RF with respect to MAGBH parameter $f_1$}\label{tab4}
    \centering 
 \scalebox{0.85}{   \begin{tabular}{c c c c c c c c c c } 
\hline                                      
        ~~~~~~~~~~~~~&$d_1=-8f_1$~~~~~~~   \\ [0.5ex]
\hline 
    $f_1$& & $r_{c}$ &~~~~~~~~~~~~~ &$v_{c}$ &  ~~~~~~~~~~~~~~~~~~~&~$H_{c}$ &~~~~~~ & \\ [0.5ex]
        \hline 
        $0.10$ ~~~~&~~   &$2.8875$&  ~~~~&~~  $0.55024$& ~~~~~&~~       $0.87174$&~~~~~~~~~~~~   \\
         $0.20$ ~~~~&~~   &$2.24675$&  ~~~~&~~  $0.84666$& ~~~~&~~         $0.65916$&~~~~~~~~~~~     \\
         $0.30$ ~~~~&~~   &$2.07655$&  ~~~~&~~  $0.47641$& ~~~~&~~        $0.60384$&~~~~~~~~~~   \\
         $0.40$ ~~~~&~~   &$1.9228$&  ~~~~&~~  $0.27785$& ~~~~&~~        $0.98581$&~~~~~~~~~~   \\
         $0.50$ ~~~~&~~   &$1.62674$&  ~~~~&~~  $0.187054$& ~~~~&~~        $0.74223$&~~~~~~~~~~~   \\
         $0.60$ ~~~~&~~   &$1.19866$&  ~~~~&~~  $0.07375$& ~~~~&~~        $0.58631$&~~~~~~~~~~~  \\
    \hline 
    \end{tabular}}
\end{table*}

\subsection{The Hamiltonian of SRF}

One can be found that the energy density is equal to four times an isotropic pressure $(4p=\rho)$.
By the relation $4f_c(r)-3r_cf'_c(r)=0$, the critical equation is given by
\begin{equation}
\frac{2\Big(5d_1k^2_s-10f_1k^2_{sh}-7mr+2r^2-20e_1k^2_d\Big)}{r^2}=0.\label{44}
\end{equation}
So, we choose only useful numerical values by the parameters $d_1=8f_1$, $f_1=1$, $e_1=1$, $m=1$ and $k_s=1$, $k_d=0.7$ $k_{sh}=2$.
\begin{equation}
r_c=0.256956.\label{45}
\end{equation}
Similarly, using the critical points, the general Hamiltonian takes the form
\begin{equation}
H=\frac{f(r)^{3/4}}{rv^{\frac{1}{2}}(1-v^{2})^{\frac{3}{4}}}.\label{46}
\end{equation}
This implies that
\begin{equation}
H=\frac{\left(1-\frac{2m}{r}+\frac{d_1k^2_s-4e_1k^2_d-2f_1k^2_{sh}}{r^2}\right)^{3/4}}{rv^{1/2}(1-v^{2})^{3/4}}.\label{46x}
\end{equation}

\begin{figure*}
\centering 
\includegraphics[width=7.2cm,height=7cm]{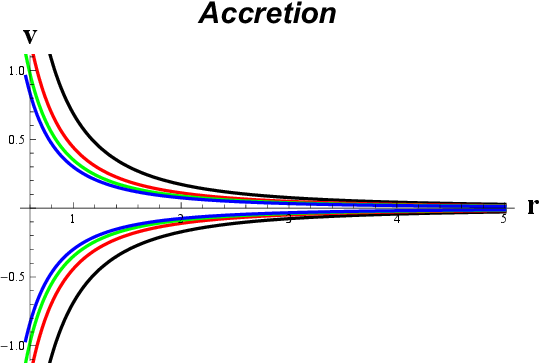}~~~
\includegraphics[width=7.2cm,height=7cm]{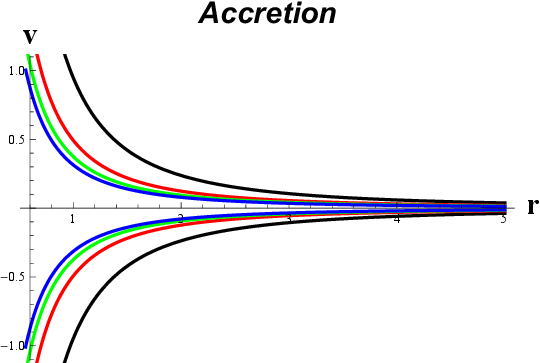}\\\hspace{2cm}
\includegraphics[width=17cm,height=6cm]{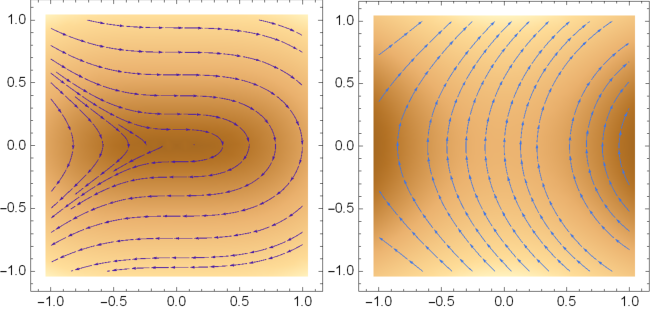}
\caption{Accretion display of stiff fluid using the MAGBH parameters. The left plot is for $d_1=8f_1$ in the lapse function $f(r)$ with the variation in $H_c$.
The right plot is for $d_1=-8f_1$ in the lapse function $f(r)$ with the variation in $H_c$. The fixed parameters are $m=1$, $e_1=0.1$, $k_s=0.85$, $k_d=0.7$ and $k_{sh}=0.7$. The bottom plots represent the streamlines accretion of stiff fluid around the MAGBH.}\label{f2}
\end{figure*}

\begin{table*}
    \caption{The sonic points for SRF with respect to MAGBH parameter $f_1$}\label{tab5}
    \centering 
  \scalebox{0.85}{  \begin{tabular}{c c c c c c c c c c }
        \hline 
 ~~~~~~~~~~~~~&$d_1=8f_1$~~~~~~~   \\ [0.5ex]
\hline 
          $f_1$& & $r_{c}$ &~~~~~~~~~~~~~ &$v_{c}$ &  ~~~~~~~~~~~~~~~~~~~&~$H_{c}$ &~~~~~~ & \\ [0.5ex]
        \hline 
         $0.10$ ~~~~&~~   &$4.15211$&  ~~~~&~~  $0.17865$& ~~~~~&~~       $0.95713$&~~~~~~~~~~~~   \\
         $0.20$ ~~~~&~~   &$3.44891$&  ~~~~&~~  $0.28675$& ~~~~&~~         $0.76952$&~~~~~~~~~~~     \\
         $0.30$ ~~~~&~~   &$3.07338$&  ~~~~&~~  $0.27214$& ~~~~&~~        $0.66359$&~~~~~~~~~~   \\
         $0.40$ ~~~~&~~   &$2.87238$&  ~~~~&~~  $0.18874$& ~~~~&~~        $0.47655$&~~~~~~~~~~   \\
         $0.50$ ~~~~&~~   &$2.38475$&  ~~~~&~~  $0.18777$& ~~~~&~~        $0.42776$&~~~~~~~~~~~   \\
         $0.60$ ~~~~&~~   &$1.29876$&  ~~~~&~~  $0.18733$& ~~~~&~~        $0.70872$&~~~~~~~~~~~  \\
    \hline 
    \end{tabular}}
\end{table*}

\begin{table*}
    \caption{The sonic points for SRF with respect to MAGBH parameter $f_1$} \label{tab6}
    \centering
   \scalebox{0.85}{ \begin{tabular}{c c c c c c c c c c } 
        \hline
 ~~~~~~~~~~~~~&$d_1=-8f_1$~~~~~~~   \\ [0.5ex]
          \hline 
    $f_1$& & $r_{c}$ &~~~~~~~~~~~~~ &$v_{c}$ &  ~~~~~~~~~~~~~~~~~~~&~$H_{c}$ &~~~~~~ & \\ [0.5ex]
        \hline 
         $0.10$ ~~~~&~~   &$2.6675$&  ~~~~&~~  $0.70711$& ~~~~~&~~       $0.45102$&~~~~~~~~~~~~   \\
         $0.20$ ~~~~&~~   &$2.44634$&  ~~~~&~~  $0.61743$& ~~~~&~~         $0.53374$&~~~~~~~~~~~     \\
         $0.30$ ~~~~&~~   &$2.12917$&  ~~~~&~~  $0.76422$& ~~~~&~~        $0.73612$&~~~~~~~~~~   \\
         $0.40$ ~~~~&~~   &$2.03375$&  ~~~~&~~  $0.70550$& ~~~~&~~        $0.75843$&~~~~~~~~~~   \\
         $0.50$ ~~~~&~~   &$1.32833$&  ~~~~&~~  $0.48709$& ~~~~&~~        $0.14497$&~~~~~~~~~~~   \\
         $0.60$ ~~~~&~~   &$1.11763$&  ~~~~&~~  $0.53515$& ~~~~&~~        $0.08771$&~~~~~~~~~~~  \\
    \hline 
    \end{tabular}}
\end{table*}

The sonic points from Tables~\ref{tab1}-\ref{tab6}, we observe that the critical radius decreased as increasing the MAGBH parameter $f_1$, but the other critical values behave differently.

\section{Results and Discussion for all test fluids}
This section provides the significance of USF, URF, RF
and SRF in the framework of MAGBH.

\subsection{Pictorial analysis of USF for MAGBH}

The accretion for the USF is depicted in Fig.~\ref{f2} for the Hamiltonian (\ref{36}). We observe from the system that there are two types of accretion flow. The first is the supersonic accretion that informed us, the fluid motion directed towards outside the horizon in the region ($v>0$). The other one is subsonic accretion, where the fluid motion is directed towards the horizon in the region ($v<0$). Both the plots (left and right) have the same analysis without the sonic points. The stream line accretion is represented in the bottom plot of Fig.~\ref{f2}. From this information, we understand that the stream lines are closed near the BH and open far from the BH. 

\subsection{Pictorial analysis of URF for MAGBH}

\begin{figure*}
\centering 
\includegraphics[width=7.2cm,height=6cm]{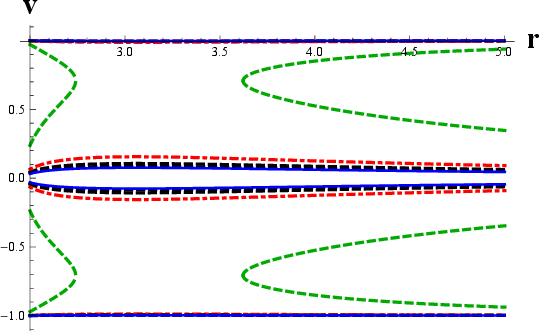}~~~
\includegraphics[width=7.2cm,height=6cm]{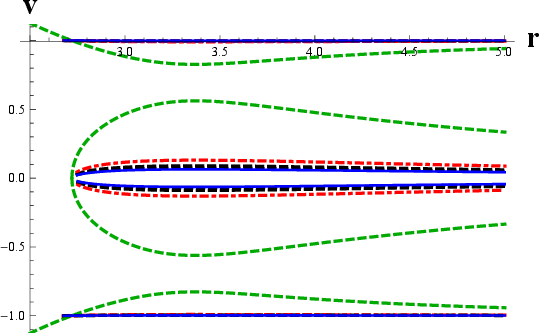}\\\hspace{2cm}
\includegraphics[width=17cm,height=6cm]{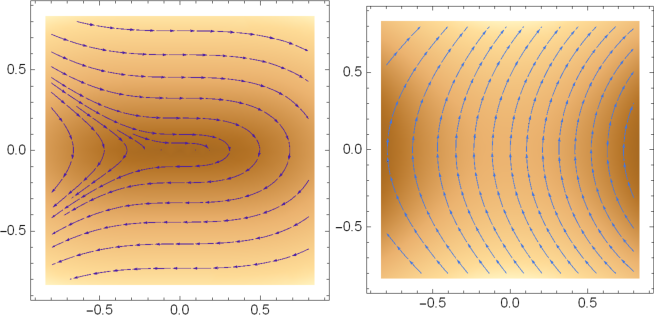}
\caption{Accretion display of ultra radiation fluid using the MAGBH parameters. Left plot is for $d_1=8f_1$ in the lapse function $f(r)$ with the variation in $H_c$. Right plot is for $d_1=-8f_1$ in the lapse function $f(r)$ with the variation in $H_c$. The fixed parameters are $m=1$, $e_1=0.1$, $k_s=0.85$, $k_d=0.7$ and $k_{sh}=0.7$. The bottom plots represent the stream lines accretion of ultra radiation fluid around the MAGBH.}\label{f3}
\end{figure*}

The accretion for the URF is depicted in Fig.~\ref{f3} for the Hamiltonian (\ref{39}).
We observe from the system that there are different types of accretion flow around the MAGBH.
The curves associated with the critical points ($r_c, v_c$) and ($r_c, -v_c$) informed us that the fluid outflow and inflow start from the event horizon, which remains very close to the BH and causes the high pressure. The purely supersonic accretion flow is observed in the neighbourhood $v<v_c$ (red dotes and blue) near the horizon, and purely subsonic accretion is observed in the neighbourhood $v>v_c$ (red dotes and blue) far from the horizon. The bottom plots in Fig.~\ref{f3} illustrate the streamlines of the accretion flow in various fluid motion patterns inside the MAGBH system.

\begin{figure*}
\centering 
\includegraphics[width=7.2cm,height=5.5cm]{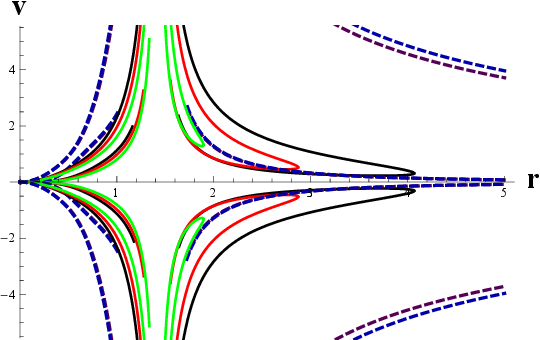}~~~~~~~
\includegraphics[width=7.2cm,height=5.5cm]{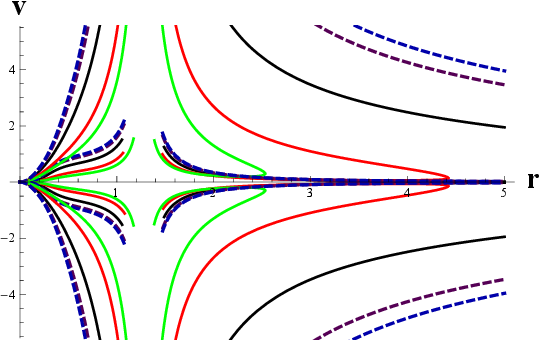}\\\hspace{2cm}
\includegraphics[width=17cm,height=6cm]{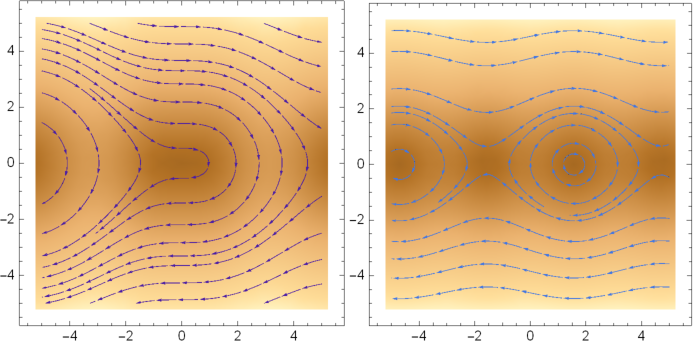}
\caption{Accretion display of radiation fluid using the MAGBH parameters. Left plot is for $d_1=8f_1$ in the lapse function $f(r)$ with the variation in $H_c$. Right plot is for $d_1=-8f_1$ in the lapse function $f(r)$ with the variation in $H_c$. The fixed parameters are $m=1$, $e_1=0.1$, $k_s=0.85$, $k_d=0.7$ and $k_{sh}=0.7$. The bottom plots represent the stream lines accretion of radiation fluid around the MAGBH.}\label{f4}
\end{figure*}

\begin{figure*}
\centering 
\includegraphics[width=7.5cm,height=5.5cm]{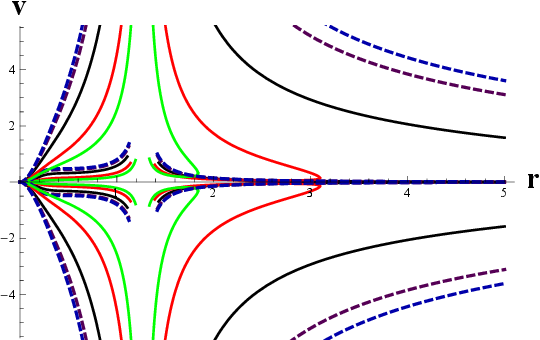}~~~~~~~~
\includegraphics[width=7.5cm,height=5.5cm]{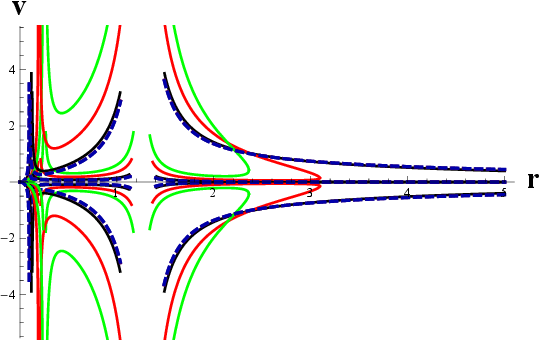}\\\hspace{2cm}
\includegraphics[width=17cm,height=6cm]{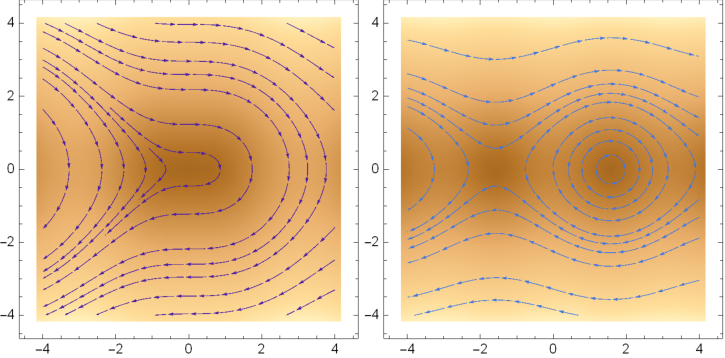}
\caption{Accretion display of sub-relativistic fluid using the MAGBH parameters. Left plot is for $d_1=8f_1$ in the lapse function $f(r)$ with the variation in $H_c$. Right plot is for $d_1=-8f_1$ in the lapse function $f(r)$ with the variation in $H_c$. The fixed parameters are $m=1$, $e_1=0.1$, $k_s=0.85$, $k_d=0.7$ and $k_{sh}=0.7$. The bottom plots represent the stream lines accretion of sub-relativistic fluid around the MAGBH.}\label{f5}
\end{figure*}

\subsection{Pictorial analysis of RF for MAGBH}

Fig.~\ref{f4} illustrates the accretion for RF with the MAGBH parameters and variation in $H_c$. We noted that the supersonic accretion of the fluid motion in the region $v>v_c$ is far from the BH for $H<H_c$ (doted curves). While the subsonic accretion of the fluid motion in the region $v<v_c$ is closest to the BH for $H>H_c$ (thick curves). We observe that the fluid motion is directed towards the BH in the form of an accretion disk at the maximum radius. The purely supersonic accretion is seen far from the source mass. The right plot of Fig.~\ref{f4}, indicates that the curves are away from the horizon for $H=H_c$ and $H>H_c$. But the supersonic and subsonic accretions are the same as seen in the left plot.
The bottom plots of Fig.~\ref{f4}, depict the stream lines accretion flow in several different types of fluid motion in the MAGBH. 

\subsection{Pictorial analysis of SRF for MAGBH}

The Fig.~\ref{f5} illustrates the accretion for SRF with the MAGBH parameters and variation in $H_c$. We noted that the supersonic accretion of the fluid motion in the region
$v>v_c$ is far from the BH for $H<H_c$ (doted curves). While the subsonic accretion of the fluid motion in the region $v<v_c$ is closest to the BH for $H>H_c$ (thick curves). We observe that the fluid motion is directed towards the BH in the form of an accretion disk at the maximum radius. The purely supersonic accretion is seen far from the source mass. The right plot of Fig.~\ref{f5}, indicates that the curves are away from the horizon for $H=H_c$ and $H>H_c$. But the supersonic and subsonic accretions are the same as seen in the left plot. The bottom plots of Fig.~\ref{f5}, depict the stream lines accretion flow in several different types of fluid motion in the MAGBH. The accurate depiction of Fig.~\ref{f5} illustrates the accretion flow in various fluid motion patterns for the MAGBH.

\section{Polytropic Fluid Accretion}

In this section, we consider the EoS for polytropic test fluid as considered by \cite{42, 43, 44}, 
\begin{equation}
p=\Gamma\, n^\gamma,\label{53}
\end{equation}
where $\Gamma$ and $\gamma$ are only constants. For an ordinary fluid, the authors considered the general
constraint $\gamma>1$. Adopting the procedure \cite{44}, we obtain the following
specific enthalpy
\begin{equation}
\chi=B+\frac{\Gamma\gamma n^{\gamma-1}}{\gamma-1},\label{54}
\end{equation}
The symbol $B$ represents the integration constant.
The determination of three-dimensional sound speed may be achieved by the use of specific enthalpy as,
\begin{equation}
\mathbf{s}^2=\frac{(\gamma-1)U}{B(\gamma-1)+U}~~~~(U\equiv \Gamma \gamma n^{\gamma-1}).\label{55}
\end{equation}
Following expression is obtained by using the speed of sound
\begin{equation}
\chi=B\frac{\gamma-1}{\gamma-1-a^2},\label{56}
\end{equation}
now we have
\begin{equation}
\chi=B\left(1+X\left(\frac{1-v^2}{r^4f(r)v^2}\right)^{(\gamma-1)/2}\right),\label{57}
\end{equation}
where
\begin{equation}
Z\equiv\frac{\Gamma \gamma n_c ^{\gamma-1}}{B(\gamma-1)}\left(\frac{r^5_c f'(r_c)}{4}\right)^{(\frac{\gamma-1}{2})}=const>0.\label{58}
\end{equation}
Here, $Z$ is a positive constant, which sustains on the BH parameters as well as on certain test fluids.
Hamiltonian can be evaluated by inserting Eq. (\ref{57}) into (\ref{31})
\begin{equation}
H=\frac{f(r)}{1-v^2}\left[1+X\left(\frac{1-v^2}{r^4f(r)v^2}\right)^{(\gamma-1)/2}\right]^2.\label{59}
\end{equation}
This implies that
\begin{small}
\begin{eqnarray}\label{60}
&&\hspace{-0.5cm}H=\frac{\left(1-\frac{2m}{r}+\frac{d_1k^2_s-4e_1k^2_d-2f_1k^2_{sh}}{r^2}\right)}{1-v^2}=\Bigg[1+X \Bigg(\frac{1-v^2}{r^4v^2\left(1-\frac{2m}{r}+\frac{d_1k^2_s-4e_1k^2_d-2f_1k^2_{sh}}{r^2}\right)}\Bigg)^{(\gamma-1)/2}\Bigg]^2.~~~~~
\end{eqnarray}
\end{small}
We examined from (\ref{60}), that $\frac{d}{dr}f(r)>0$, for all $r$.
However, the coefficient of $\frac{f(r)}{1-v^2}$, diverges when $r\rightarrow\infty$,
and the denominator $1-v^2$, assume the domain $(0, 1)$, hence the Hamiltonian diverges.

Following the procedure \cite{42, 43, 44}, we obtain the following relations
\begin{equation}\label{63}
(\gamma-1-v^2_c)\left(\frac{1-v^2_c}{r^4_cf(r_c)v^2_c}\right)^{\frac{\gamma-1}{2}}=\frac{n_c}{2X}\left(r^5_cf'(r_c)\right)^{\frac{1}{2}v^2_c},
\end{equation}
\begin{equation}\label{64}
v^2_c = \frac{r_cf'_{r_c}}{r_cf'_{r_c}+4f(r_c)}.
\end{equation}

\begin{figure*}
\centering 
\includegraphics[width=7.5cm,height=5.5cm]{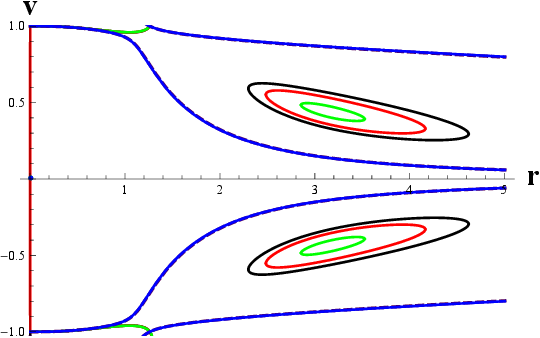}~~~~~~~~
\includegraphics[width=7.5cm,height=5.5cm]{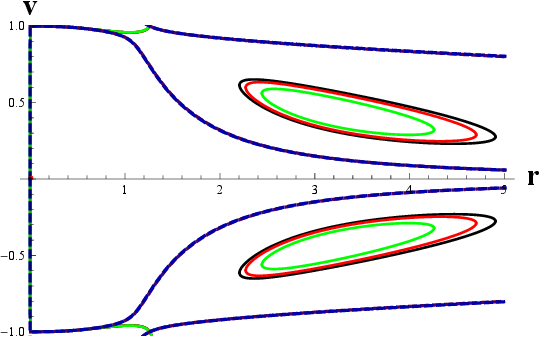}\\\hspace{2cm}
\includegraphics[width=17cm,height=6cm]{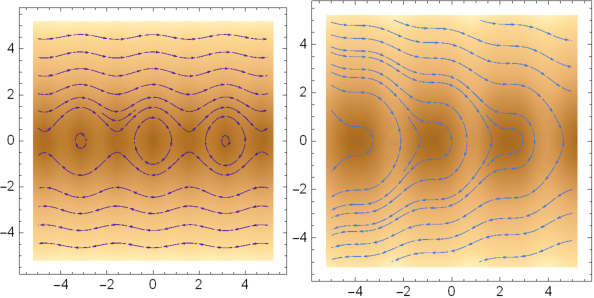}
\caption{Accretion display of polytropic test fluid using the MAGBH parameters. Left plot is for $d_1=8f_1$ in the lapse function $f(r)$ with the variation in $H_c$. Right plot is for $d_1=-8f_1$ in the lapse function $f(r)$ with the variation in $H_c$. The fixed parameters are $m=1$, $e_1=0.1$, $k_s=0.85$, $k_d=0.7$ and $k_{sh}=0.7$. The bottom plots represent the stream lines accretion of polytropic test fluid around the MAGBH.}\label{f6}
\end{figure*}

The accretion of polytropic test fluid around the MAGBH with extra constants $B=0.15$, $Z=7$, $\gamma=5/3$ is depicted in Fig.~\ref{f6}. This analysis is important because it gives us a better understanding of the circular disks produced by the accreting BH. We observe that the critical flow begins on a subsonic path, then follows a supersonic path, escaping the sonic point, and ends at the killing horizon. As the critical flow towards the BH, it develops into the accretion disks. These disks are below the critical velocities and at the maximum radius. From the bottom contour, there are the same types of accretion flow for polytropic test fluid with parameter $\gamma$ in the stream lines.

\section{Mass Accretion Rate Significance}

Since the mass of a BHs as well as a dark compact object is increased by the accreting matter falling onto them. The mass rate of change or accretion rate of a MAGBH dark compact object can be explored through $\dot{M}\mid_{rh}=4\pi r^2 T^r_t \mid_{rh}$ \cite{57}.
For this, we consider $T^r_t=(\rho+p)u_t u^r$ \cite{57rr,57rrr} and the conserved elements
$\Delta_\eta J^\eta=0$ and $\nabla_\zeta T^{\eta\zeta}=0$.
Using the Eqs. (\ref{13}) and (\ref{17}), we get
\begin{equation}
r^2u(\rho+p)\Big(f(r)+(u)^2\Big)^{1/2}=A_0,\label{65}
\end{equation}
where $A_0$ represents an arbitrary constant. The energy flow and the equation of state, denoted as $p=p(\rho)$, lead us to 
\begin{equation}
\frac{1}{\rho+p}d\rho+\frac{1}{u}du+\frac{2}{r}dr=0.\label{66}
\end{equation}
Integrating Eq. (\ref{66}), we get
\begin{equation}
\exp\left[\int^{\rho}_{\rho\infty}\frac{d \rho'}{\rho'+p(\rho')}\right]ru =-A_1,\label{67}
\end{equation}
where $A_1$ is an integration constant and $\rho_\infty$ shows a fluid density at infinity.
The Eqs. (\ref{65}) and (\ref{67}), yield

\begin{eqnarray}
&& \hspace{-0.5cm}A_3=-\frac{A_0}{A_1}=(\rho+p)\Big(f(r)+(u)^2\Big)^{1/2} \exp\left[-\int^{\rho}_{\rho\infty}\frac{d \rho'}{\rho'+p(\rho')}\right],\label{68}
\end{eqnarray}

where $A_3$ is constant of integration. While, $A_3=\rho_\infty +p(\rho_\infty)=-\frac{A_0}{A_1}$,
with $A_0=(\rho+p)u^0 u r^2=-A_1(\rho_\infty +p(\rho_\infty))$ at infinity.
The given problem is static and spherically symmetric at equatorial plane, so, $\nabla_\eta J^\eta=0$, we get
\begin{equation}
r^2u\varepsilon=A_2,\label{69}
\end{equation}
where $A_2$ is constant of integration.
The Eqs. (\ref{65}) and (\ref{69}), give the relation
\begin{equation}
\frac{\rho+p}{\varepsilon}\Big(f(r)+(u)^2\Big)^{1/2}=\frac{A_0}{A_2}\equiv A_4,\label{70}
\end{equation}
and $A_4$ is any constant as $A_4=\frac{(\rho_\infty+p_\infty)}{\varepsilon_\infty}$.
Keeping the Eq. (\ref{65}), the mass of BH takes the form
\begin{equation}
\dot{M}=-4\pi r^2u(\rho+p)\Big(f(r)+(u)^2\Big)^{1/2}=-4\pi A_0.\label{71}
\end{equation}
This implies that
\begin{equation}
\dot{M}=4\pi A_1(\rho_\infty+p(\rho_\infty)).\label{72}
\end{equation}
This Eq. (\ref{72}) yields the valid outcome for any nature of fluid.
Therefore
\begin{equation}
\dot{M}=4\pi A_1(\rho+p)|_{r=r_h},\label{73}
\end{equation}
Using EoS $(p=\omega\rho)$, into the Eq. (\ref{67}), we get
\begin{equation}
\rho=\left[-\frac{A_1}{r^2u}\right]^{1+\omega}.\label{74}
\end{equation}
Now using this expression into (\ref{65}), we obtain
\begin{equation}
(u)^2-\frac{A^2_0A^{-2(1+\omega)}_1}{(1+\omega)^2}r^{4\omega}(-u)^{2\omega}+f(r)=0.\label{75}
\end{equation}
This result is valid for fluid velocity $u$ for any value of $\omega$.
Simply, one can be found the energy density $\rho$ by using this result.

\subsection{Mass Accretion Rate for USF}
One can find the radial velocity and the energy-density
keeping $\omega=1$ in Eqs. (\ref{70}) and (\ref{74}), we have
\begin{equation}
u=\pm A_1^2\sqrt{\frac{f(r)}{A^2_0r^4-4A^4_1}},\label{76}
\end{equation}
and
\begin{equation}
\rho=\frac{(A^2_0r^4-4A^4_1)}{4A^2_1r^4f(r)}.\label{77}
\end{equation}

Thus, from (\ref{73}) and (\ref{77}), we evaluate the mass accretion rate for
MAGBH, which is given by
\begin{eqnarray}\label{78}
\dot{M}&=&\frac{2\pi (A^2_0r^4-4A^4_1)}{A_1r^4f(r)}.\\
\dot{M}&=&\frac{2\pi (A^2_0r^4-4A^4_1)}{A_1r^4\Big(1-\frac{2m}{r}+\frac{d_1k^2_s-4e_1k^2_d-2f_1k^2_{sh}}{r^2}\Big)}\label{78a}.
\end{eqnarray}
\begin{figure*}
\centering
\includegraphics[width=80mm,height=6cm]{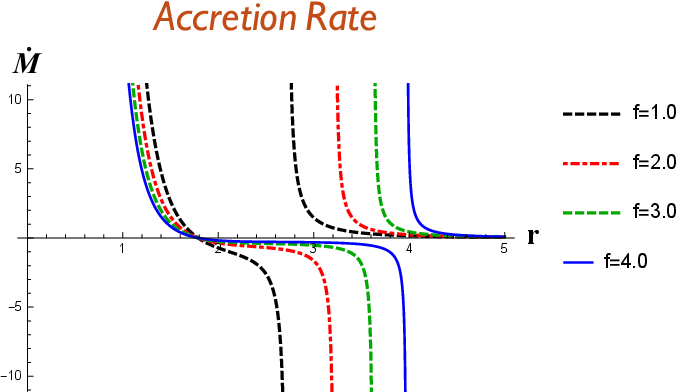}~~~~~
\includegraphics[width=80mm,height=6cm]{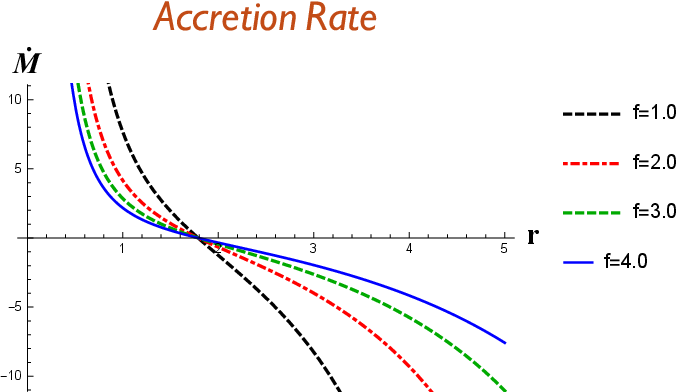}
\caption{Accretion rate display of stiff fluid using the MAGBH parameters. Left plot is for $d_1=8f_1$ in the lapse function $f(r)$ with the variation in $f_1$. Right plot is for $d_1=-8f_1$ in the lapse function $f(r)$ with the variation in $f_1$. The fixed parameters are $m=1$, $e_1=0.1$, $k_s=0.85$, $k_d=0.7$ and $k_{sh}=0.7$.}\label{f7}

\end{figure*}

\begin{figure*}
\begin{center}
\includegraphics[width=80mm,height=6cm]{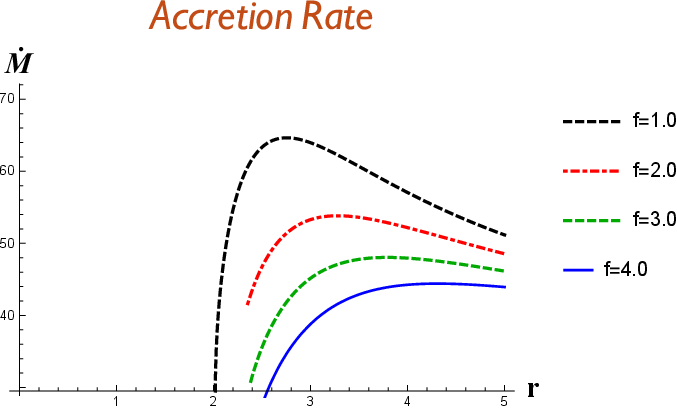}~~~~~
\includegraphics[width=80mm,height=6cm]{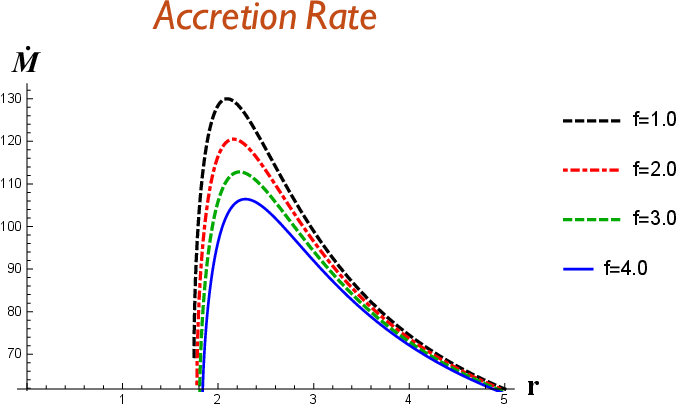}
\caption{Accretion rate display of ultra radiation fluid using the MAGBH parameters. Left plot is for $d_1=8f_1$ in the lapse function $f(r)$ with the variation in $f_1$. Right plot is for $d_1=-8f_1$ in the lapse function $f(r)$ with the variation in $f_1$. The fixed parameters are $m=1$, $e_1=0.1$, $k_s=0.85$, $k_d=0.7$ and $k_{sh}=0.7$.}\label{f8}
\end{center}
\end{figure*}

\begin{figure*}
\begin{center}
\includegraphics[width=80mm,height=5.5cm]{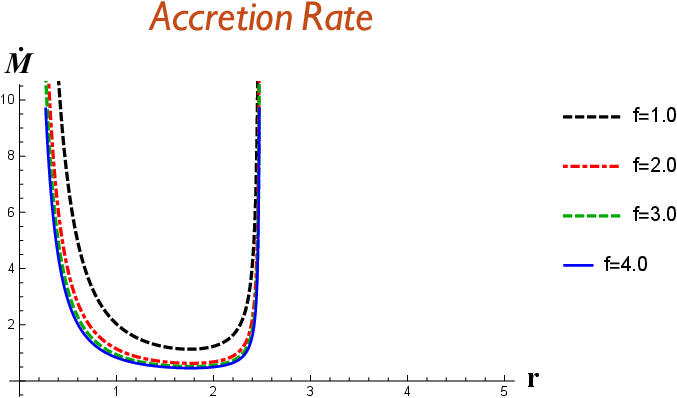}~~~~~
\includegraphics[width=80mm,height=5.5cm]{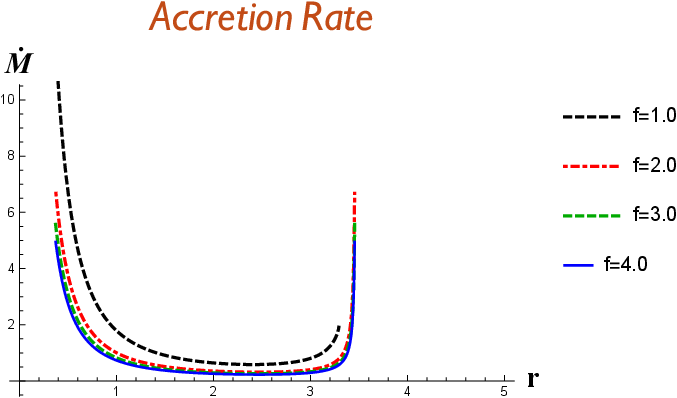}
\caption{Accretion rate display of radiation fluid using the MAGBH parameters. Left plot is for $d_1=8f_1$ in the lapse function $f(r)$ with the variation in $f_1$. Right plot is for $d_1=-8f_1$ in the lapse function $f(r)$ with the variation in $f_1$. The fixed parameters are $m=1$, $e_1=0.1$, $k_s=0.85$, $k_d=0.7$ and $k_{sh}=0.7$.}\label{f9}
\end{center}
\end{figure*}

\begin{figure*}
\begin{center}
\includegraphics[width=80mm,height=6cm]{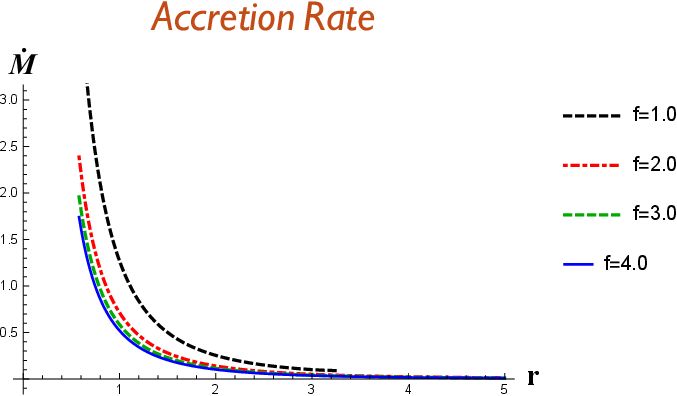}~~~~~
\includegraphics[width=80mm,height=6cm]{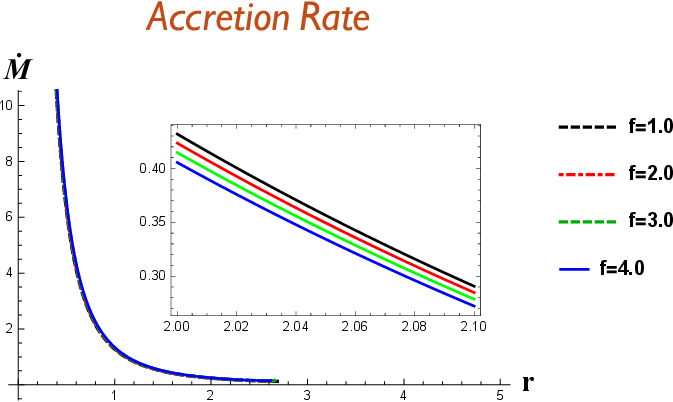}
\caption{Accretion rate display of sub-relativistic fluid using the MAGBH parameters. Left plot is for $d_1=8f_1$ in the lapse function $f(r)$ with the variation in $f_1$. Right plot is for $d_1=-8f_1$ in the lapse function $f(r)$ with the variation in $f_1$. The fixed parameters are $m=1$, $e_1=0.1$, $k_s=0.85$, $k_d=0.7$ and $k_{sh}=0.7$.}\label{f10}
\end{center}
\end{figure*}
The accretion rate $\dot{M}$ versus the radius $r$
for USF around the MAGBH is analyzed in Fig.~\ref{f7}.
In the left plot, the accretion rate remains the same at the maximum position $(\dot{M}=10)$ for distant observers. The radius increased as increasing the parameter $f_1$.
In the right plot, the same behavior is observed as in the left plot, but the observer position is fixed at $r=1.91$.

\subsection{Mass Accretion rate for URF}
One can find the radial velocity and the energy density
keeping $\omega=1/2$ in Eqs. (\ref{70}) and (\ref{74}), we have,

\begin{small}
\begin{eqnarray}
&& \hspace{-0.5cm}u=\frac{2r^2A^2_0+\sqrt{4r^2A^4_0-81f(r)A^6_1}}{9A^3_1},\label{81}\\
&& \hspace{-0.5cm}\rho=27\left(\frac{A^4_1}{r^2\Big(2r^2A^2_0+\sqrt{4r^2A^4_0-81f(r)A^6_1}\Big)}\right)^{3/2},\label{82}\\
&& \hspace{-0.5cm}\dot{M}=216\pi A_1\left(\frac{A^4_1}{r^2\Big(2r^2A^2_0+\sqrt{4r^2A^4_0-81f(r)A^6_1}\Big)}\right)^{3/2}.\label{83}
\end{eqnarray}
\end{small}
This implies that
\begin{eqnarray}
\dot{M}=216\pi A_1\left(\frac{A^4_1}{r^2\Big(2r^2A^2_0+M_{1}\Big)}\right)^{3/2},\label{84}
\end{eqnarray}
where, 
\begin{small}
\begin{eqnarray}
   M_1=\sqrt{4r^2A^4_0-81\Big(1-\frac{2m}{r}+\frac{d_1k^2_s-4e_1k^2_d-2f_1k^2_{sh}}{r^2}\Big)A^6_1}. \nonumber 
\end{eqnarray}
\end{small}

\subsection{Mass Accretion rate for RF}
One can find the radial velocity and the energy density
keeping $\omega=1/3$ in Eqs. (\ref{70}) and (\ref{74}), we have
\begin{small}
\begin{eqnarray}
&& \hspace{-0.4cm} u=\Big[\frac{\Big(-32f(r)A_1^4+\Psi_{11}(r)\Psi_{11}(r)\Big)^{1/3}}{4A_1^2}
+\frac{3r^{4/3}A_0^2}{4A_1^{2/3}\Big(\Psi_{11}(r)-32f(r)A_1^4\Big)^{1/3}}\Big]^{2/3},~~~~~~\label{87}
\end{eqnarray}
\end{small}
and
\begin{small}
\begin{eqnarray}
&& \hspace{-0.4cm}\rho=\Big[\frac{A_1}{r^2}\Big]^{\frac{4}{3}}\Big[\frac{\Big(-32f(r)A_1^4+\Psi_{11}(r)\Big)^{1/3}}{4A_1^2}
+\frac{3r^{4/3}A_0^2}{4A_1^{2/3}\Big(\Psi_{11}(r)-32f(r)A_1^4\Big)^{1/3}}\Big]^{\frac{-8}{9}}.~~~~~~\label{88}
\end{eqnarray}
\end{small}
where, $\Psi_{11}(r)=\sqrt{1024f(r)^2A_1^8-27r^4A_0^6}$.
The mass accretion rate of BH is given by

\begin{eqnarray}\label{89}
 \dot{M}=\Big[\frac{8\pi A_1^{\frac{7}{3}}}{r^{\frac{8}{3}}}\Big]\Big[\Big(\Psi_{11}(r)-32f(r)A_1^4\Big)^{1/3} \Big({4A_1^2}\Big)^{-1}+\frac{3r^{4/3}A_0^2}{4A_1^{2/3}}
\Big(\Psi_{11}(r)-32f(r)A_1^4\Big)^{-1/3}\Big]^{\frac{-8}{9}}.
\end{eqnarray}

This implies that
\begin{small}
\begin{eqnarray}\label{90}
&& \hspace{-0.5cm}\dot{M}=\Big[\frac{8\pi A_1^{\frac{7}{3}}}{r^{\frac{8}{3}}}\Big]\Big[\Big(-32\Big(1-\frac{2m}{r}+\frac{d_1k^2_s-4e_1k^2_d-2f_1k^2_{sh}}{r^2}\Big)A_1^4
+\sqrt{M_2-27r^4A_0^6}\Big)^{1/3}\Big({4A_1^2}\Big)^{-1}
\nonumber\\&&+\frac{3r^{4/3}A_0^2}{4A_1^{2/3}}\Big(-32\Big(1-\frac{2m}{r}+\frac{d_1k^2_s-4e_1k^2_d-2f_1k^2_{sh}}{r^2}\Big)A_1^4 +\sqrt{M_2-27r^4L_0^6}\Big)^{-1/3}\Big]^{\frac{-8}{9}}.
\end{eqnarray}
\end{small}
where,
\begin{small}
\begin{eqnarray}
    M_2=1024\Big(1-\frac{2m}{r}+\frac{d_1k^2_s-4e_1k^2_d-2f_1k^2_{sh}}{r^2}\Big)^2A_1^8. \nonumber
\end{eqnarray}
\end{small}

\subsection{Mass Accretion rate for SRF}

Herein this subsection, we shall find the radial velocity and the energy density
by considering $\omega=1/4$ in Eqs. (\ref{70}) and (\ref{74}), one can get the following expression for radial velocity:
\begin{small}
\begin{eqnarray}
&& \hspace{-0.4cm} u=\frac{\sqrt{u_2}}{5 \sqrt{3}}-\frac{1}{2} \bigg(-\frac{256 \sqrt{3} A_0^4 r^2}{125 B^5 \sqrt{u_2}}-\frac{100\ 5^{2/3} A_1^5 f^2}{3 \sqrt[3]{u_1}}-\frac{4 \sqrt[3]{u_1}}{75\ 5^{2/3} A_1^5}-\frac{8 f}{3}\bigg)^{\frac{1}{2}}~~~~~~\label{91}
\end{eqnarray}
\end{small}
where 
\begin{small}
\begin{eqnarray*}
u_1&&=13824 A_0^8 A_1^5 r^4+96 \sqrt{3} \bigg(6912 A_0^{16} A_1^{10} r^8-390625 A_0^8 A_1^{20} f^3 r^4\bigg)^{\frac{1}{2}}-390625 A_1^{15} f^3,\\
u_2
&&=\frac{625\ 5^{2/3} A_1^5 f^2}{\sqrt[3]{u_1}}+\frac{\sqrt[3]{u_1}}{5^{2/3} A_1^5}-25 f.
\end{eqnarray*}
\end{small}
The energy density for SRF fluid is calculated as:
\begin{small}
\begin{eqnarray}
&& \hspace{-0.4cm}\rho=\left(-\frac{A_1}{r^2 \left(\frac{\sqrt{\frac{u_2 \sqrt[3]{u_1}}{5^{2/3} A_1^5}}}{5 \sqrt{3}}-\frac{1}{2} \sqrt{\rho _1-\frac{256 \sqrt{3} A_0^4 r^2}{125 A_1^5 \sqrt{\frac{u_2 \sqrt[3]{u_1}}{5^{2/3} A_1^5}}}}\right)}\right)^{5/4},~~~~~~\label{92}
\end{eqnarray}
\end{small}
where
\begin{small}
\begin{eqnarray}\label{93}
&& \hspace{-0.5cm}\rho _1=-\frac{100\ 5^{2/3} A_1^5 f^2}{3 \sqrt[3]{u_1}}-\frac{4 \sqrt[3]{u_1}}{75\ 5^{2/3} A_1^5}-\frac{8 f}{3}.
\end{eqnarray}
\end{small}
Finally, acceration rate is calculated as
\begin{small}
\begin{eqnarray}\label{94}
&& \hspace{-0.5cm}\dot{M}=5 \pi  A_1 \left(-\frac{A_1}{r^2 \left(\frac{\sqrt{\frac{u_2 \sqrt[3]{u_1}}{5^{2/3} A_1^5}}}{5 \sqrt{3}}-\frac{1}{2} \sqrt{\rho _1-\rho _2}\right)}\right)^{5/4},
\end{eqnarray}
\end{small}
where 
\begin{small}
\begin{eqnarray}\label{98}
&& \hspace{-0.5cm}\rho _2=\frac{256 \sqrt{3} A_0^4 r^2}{125 A_1^5 \sqrt{\frac{u_2 \sqrt[3]{u_1}}{5^{2/3} A_1^5}}}.
\end{eqnarray}
\end{small}
Fig.~\ref{f8} illustrates the accretion rate for URF near the MAGBH. In both cases, we observe that the accretion rate decreased as the parameter $f_1$ increased and vice versa. In Fig.~\ref{f9}, it is decreased towards the BH and increased far from the BH for RF as increasing values of $f_1$. For SRF in Fig.~\ref{f10}, it is a decreasing function of $r$ as increasing the parameter $f_1$.

\section{Conclusion}

Within the context of the traceless nonmetricity tensor, we have examined the MAG in the current study, which is predicated on the metricity of space-time and the affine relationship. Stated differently, it illustrates the relationship between the metric tensor and the affine connection derived from the Levi-Civita connection. In fact, MAG, a theory of gravity that extends GR by permitting non-zero torsion and nonmetricity, heavily relies on the traceless nonmetricity tensor. It is interesting to note that these extra geometric objects are related, respectively, to the spin of matter fields and the existence of a preferred direction in space-time. The presence of nonmetricity assists in the computation of the geodesic equation for test particles, resulting in departures from the GR-predicted acceleration. Further, we have explored some interesting aspects of matter accretion onto charged BH solutions in the framework of MAG. For this purpose, we have discussed general accretion by using different kinds of fluids. Further, we have used the Hamiltonian approach to discuss the accretion in the current study. Some important insights from the present analysis are listed below: 
\begin{itemize}
\item As we have discussed Reissner-Nordstr$\ddot{o}$m-like BH geometry with spin, dilation, and shear charges. In the considered BH solution, we have two different cases, i.e., $d_1 = 8f_1$, (ii) $d_1=-8f_1$ and (iii) $d_1\neq8f_1$ for which $f_1 \leq 0$. The horizon structure for the considered BH solution is shown in Fig.~\ref{f1}. It has been noted that singularities exist in both cases of the BH solution.  
\item We have developed the fundamental formulas for general accretion to exploring the critical flow of fluid in the background of spherically symmetric MAGBH. We have also provided the basic background for sonic points and isothermal test fluid.  
\item Four different kinds of fluids as, USF ($\omega=1$), URF ($\omega=1/2$), RF ($\omega=1/3$) and SRF ($\omega=1/4$) have been considered in the accretion flow around a MAGBH. It is necessary to mention that we have successfully calculated these four different kinds of fluids for the considered MAGBH solution.

\item For the Hamiltonian through Eq. (\ref{36}), the accretion for the USF is shown in Fig.~\ref{f2}. Two different forms of accretion flow are visible in the system. The first one is the fluid motion oriented towards the outside of the horizon in the region ($v>0$), which is what the supersonic accretion told us about. An alternative scenario involves subsonic accretion, in which fluid motion is directed towards the horizon within the region ($v<0$). The bottom plot of Fig.~\ref{f2} shows the accretion through stream lines. From these stream lines, we are able to understand that the stream lines are closed near the BH and they are open far from the BH.

\item The URF's accretion is given in Fig.~\ref{f3}. We can see from the system that the accretion flow surrounding the MAGBH is diverse. The critical point curves ($r_c, v_c$), ($r_c, -v_c$) showed us that the high pressure is caused by the fluid outflow and inflow starting at the event horizon and continuing quite close to the BH. Several different kinds of fluid motion in the MAGBH are depicted in the bottom plots of Fig.~\ref{f3}.

\item The accretion for RF with the MAGBH settings and change in $H_c$ is shown in Fig.~\ref{f4}. For $H<H_c$, we observed that the supersonic accretion of the fluid motion is located far from the BH in the area $v>v_c$ (dot curves). For $H>H_c$ (thick curves), the subsonic accretion of the fluid motion is closest to the BH in the region $v<v_c$. We note that the fluid motion at the maximum radius is directed towards the BH as an accretion disk. Fig. (\ref{f4}) bottom plots show the stream lines accretion flow in various forms of fluid motion in the MAGBH.

\item The accretion for SRF with the MAGBH parameters and change in $H_c$ are provided in Fig.~\ref{f5}. We note that the fluid motion at maximum radius is directed towards the BH as an accretion disk. The accretion is completely supersonic and observed far from the source mass. For both $H=H_c$ and $H>H_c$, the curves are away from the horizon, according to the right plot of same figure. However, from the left plot, it has been confirmed that the supersonic and subsonic accretions are the same. Fig.~\ref{f5} bottom plots show the stream lines accretion flow in various forms of fluid motion in the MAGBH.

\item The Fig.~\ref{f6} shows the accretion for polytropic test fluid around the MAGBH with the introduced constants $B=0.15$, $Z=7$, and $\gamma=5/3$. This study is significant because it helps us comprehend the circular disks that the accreting BH produces. As we can see, the crucial flow starts on a subsonic course, escapes the sonic point by taking a supersonic path, and finishes at the killing horizon. The accretion disks form when the critical flow moves closer to the BH. These disks are at the maximum radius and below the critical velocities.

\item The mass accretion rates for USF, URF, RF, and SRF in the background of the MAGBH solution are provided graphically in the Figs. \ref{f7}-\ref{f10}, respectively. For distant observers, the accretion rate is constant at the maximum position $(\dot{M}=10)$. As the parameter $f_1$ increased, the radius also increased. One can see similar behavior in the right plot for USF. In both scenarios, we see that when we increased the parameter $f_1$, the accretion rate fell within the context of URF, and vice versa. For RF, it decreases near the BH and increases further from the BH as $f_1$ values grow. In the SRF fluid, the accretion rate decreases as the parameter $f_1$ increases.
\end{itemize}
The study has examined the effects of parameter $f_1$ on a MAGBH's horizon structure. Within the domain of accretion, multiple fluids have been examined depending on the value of $f_1$. Using a variety of models of fluids, including the USF, URF, RF, and SRF, the behavior of accretion flow has been investigated. The identification of critical points and flow patterns revealed supersonic and subsonic accretion in various places. The investigation produced well-behaved and physically realistic outcomes. These studies provide a new perspective regarding the accretion of BH.

\end{document}